\documentstyle[manuscript,aps]{revtex}

\input epsf

\def\beq{\begin{equation}}
\def\eeq{\end{equation}}
\def\beqn{\begin{eqnarray}}
\def\eeqn{\end{eqnarray}}
\def\bsigma{\mbox{\boldmath $\sigma$}}
\def\balpha{\mbox{\boldmath $\alpha$}}
\def\bT{{\bf T}}
\def\bx{{\bf x}}
\def\by{{\bf y}}
\def\bs{{\bf s}}
\def\br{{\bf r}}
\def\bz{{\bf z}}

\begin{document}
\draft

\preprint{CU-TP-881, {\tt hep-th}/9802108}
%\

\setcounter{page}{0}

\title{\Large\bf SU(2) Calorons and Magnetic Monopoles}

\author{\large\it
Kimyeong Lee\footnote{Electronic address: klee@phys.columbia.edu}
and Changhai Lu\footnote{Electronic address: chlu@cuphy3.phys.columbia.edu}}

\address{Physics Department, Columbia University, New York, NY, 10027}
\date{\today}

\maketitle

\begin{abstract}
We investigate the self-dual Yang-Mills gauge configurations on
$R^3\times S^1$ when the gauge symmetry SU(2) is broken to U(1) by
the Wilson loop.  We construct the explicit field configuration for a
single instanton by the Nahm method and show that an instanton is
composed of two self-dual monopoles of opposite magnetic charge.
We  normalize the moduli space metric of an instanton and study
various limits of the field configuration  and its moduli space metric.
\end{abstract}

\vspace{2mm}


\setcounter{footnote}{0}

\section{Introduction}

Recently there has been some interest in understanding the relation
between calorons, or periodic instantons, and magnetic monopoles on
$R^3\times S^1$, where the gauge symmetry is broken maximally by the
Wilson loop. Especially it has been argued that instantons are
composite objects made of magnetic
monopoles~\cite{piljin,kimyeong,nahm}.  Among self-dual configurations
of a theory with a simple gauge group $G$ of rank $r$, the
configurations independent of $S^1$-coordinate satisfy the ordinary BPS
equations for the magnetic monopoles.  On $R^3$ there exist $r$ types
of fundamental BPS magnetic monopoles associated with each simple
roots~\cite{erick}.  On $R^3\times S^1$, it was shown that there
exists an additional type of fundamental monopoles associated with the
lowest negative root~\cite{piljin,kimyeong}.  It was argued that a
single instanton is made of unique combination of $r+1$ different
fundamental monopoles such that the net magnetic charge is
zero~\cite{kimyeong,piljin}. Also the explicit moduli space metric of
a single instanton in the $SU(n)$ theory has been obtained up to
normalization~\cite{piljin}.

In this paper we construct the explicit field configuration for a
single $SU(2)$ caloron on $R^3\times S^1$ with nontrivial Wilson loop
by using the Nahm construction and show that a single caloron is
made of two distinct fundamental magnetic monopoles. We also examine
various limits of the configuration, especially the trivial Wilson
loop limit and the zero temperature limit.  We also investigate the
moduli space and its metric.

For convenience, we imagine a five dimensional theory with additional
time direction $x_0$. Thus our instantons and monopoles will appear as
solitons in this theory. However, they may also play an important role
in finite temperature Yang-Mills theory where $x_4$ plays the
Euclidean time.  The caloron, or periodic instanton, solutions have
been found late seventies~\cite{caloron,rossi,pisarski}.  The
difference between those works and ours lies on the Wilson loop
$W(\bx)= P\exp (\int dx_4 A_4) $. In all those cases~\cite{caloron}
the Wilson loop is trivial and so magnetic monopole solutions appear
only when the scale of the instanton is taken to be
infinity~\cite{rossi}. In our case, the Wilson loop is nontrivial.  In
a chosen gauge the value of $A_4$ at spatial infinity
\beq
<A_4>=  -i \frac{u}{2} \sigma_3
\label{back}
\eeq
plays the  role of  Higgs expectation value.

In the Feynman path integral, we can require that only the field
configurations periodic in $x_4\in [0,\beta]$ contribute.  The
allowed local gauge transformations are the ones which leave the gauge
field  single-valued.  For the gauge group $SU(2)$, there is a
group of topologically nontrivial (large) gauge transformations, for
example,
\beq
U_L(x_4) =
\exp ( -i\frac{\pi x_4}{\beta} \sigma_3 ).
\label{large}
\eeq
Even though it is not single-valued as $U_L(x_4+\beta)= - U_L(x_4)$,
it is acceptable since  the gauge fields remain single-valued.  Using this
large gauge transformation and the Weyl reflection $e^{i\frac{\pi}{4}
\sigma_2}$, which sends $u\rightarrow -u$, we can choose the
range of $u$ to be
\beq
 0\le u \le \frac{2\pi}{\beta}.
\label{weyl}
\eeq 
When $u\ne 0, 2\pi/\beta$, one can see easily that the gauge symmetry
to be spontaneously broken from $SU(2)$ to $U(1)$. There is also an
additional global $U(1)$ symmetry corresponding to the translational
symmetry on $S^1$~\cite{piljin}.  (Of course, one can gauge away the
background field (\ref{back}) once we impose the condition
$A_\mu(x_4+\beta) = e^{i\frac{u}{2}\sigma_3} A_\mu(x_4)
e^{-i\frac{u}{2}\sigma_3} $ for acceptable gauge configurations.)
 
In the normalization where the coupling constant $e^2=1$, the action,
or four dimensional energy, is bounded from below, $S \ge 8\pi^2 |k|$,
by the topological index
\beqn k &=&\frac{1}{64\pi^2}\int d^4x \, \epsilon_{\mu\nu\rho\sigma}
F_{\mu\nu}^a F^a_{\rho\sigma} \nonumber\\ 
&=& \frac{1}{16\pi^2}
 \int d^3S_i \,\epsilon_{ijk} (F_{ij}^a A_4^a - A^a_j\partial_4 A^a_k).
\label{top}
\eeqn
The  boundary contributions can be nonzero  near  gauge singularities and 
spatial infinity. When $k>0$, the bound is saturated by the field
configurations satisfying  self-dual equations 
\beq
F_{ij} = \epsilon_{ijk} (D_k A_4-\partial_4 A_k). 
\eeq

When the asymptotic value of $A_4$ lies in the interval (\ref{weyl}),
it was shown that there exist self-dual configurations for two kinds
of fundamental magnetic monopoles of four zero
modes~\cite{piljin,kimyeong}. One configuration is the ordinary BPS
solution, which is independent of $x_4$. It describes monopoles of
positive magnetic charge $4\pi$ and asymptotic Higgs value
$u$. Another solution is an ordinary monopole with asymptotic Higgs
value $2\pi/\beta -u$. We need to apply a large gauge transformation
(\ref{large}) and a Weyl reflection to this solution to get the right
boundary condition.  It describes monopoles of negative magnetic
charge $-4\pi$. The topological charges of
both type of monopoles are positive and are given, respectively, by
\beq
k_1 = \frac{\beta u }{2\pi},\,\, k_2=1-\frac{\beta u}{2\pi}
\eeq
The masses of  magnetic monopoles in conventional sense are  the
magnetic charge times the length scale, and so
\beq
m_1 =4\pi u , \, m_2=4\pi (\frac{2\pi}{\beta} -u)
\eeq
As five dimensional solitons, the monopoles really carry mass $\beta
m_1$ and $\beta m_2$. Each type of monopoles can carry electric charge
$q_i$, which is integer quantized as they arise from $W$ boson
excitations.

The reason for the opposite charge of these two monopoles can be seen
easily in the unitary gauge. For the first monopole, $A_4$ increases
from zero to $u$ as one moves from monopole core to spatial
infinity. For the second monopole, the value of $A_4$ decreases from
$2\pi/\beta$ to $u$ as one moves from monopole core to spatial
infinity. The magnetic field is proportional to the spatial derivative
of $A_4$ and so the two monopoles carry opposite charge. However,
there is no static force between them because the Higgs interaction is
now repulsive, as one can see from the mass formula, and it cancels
the magnetic attraction exactly. That is why in principle two solutions
can be superposed. The configurations for superposed two distinct
fundamental monopoles will satisfy the self-dual equations and have
zero total magnetic charge, unit topological charge, and eight zero
modes. Those are exactly the field configurations for a single
instanton.

Another interesting question is to find the moduli space metric.  The
moduli space of a single instanton on $R^3 \times S^1$ is found up to
right coefficients. Especially, the relative moduli space
for a single instanton was argued to be ${\rm Taub-NUT}$ with $Z_2$
singularity~\cite{piljin}. We find the exact moduli space metric and
the moduli space by using the constituent monopole
picture~\cite{lwy2,lwy1}.

The plan of this paper is as follows. In Sec.~II, we briefly review
the Nahm formalism and use it to construct the field configuration for
a single instanton on $R^3\times S^1$ with the nontrivial Wilson loop.
In Sec.~III, we show that the field configuration approaches the
single monopole configuration at the expected positions of
monopoles. This shows that a single instanton solution is a
complicated superposition of two monopole configurations. In Sec.~IV,
we study the field configuration outside monopole core region. In
Sec.~V, we show that our solution has a gauge singularity at
center-of-mass, which leads to the unit topological charge.  In
Sec.~VI, we take the limit where one of the monopoles becomes massless
and show that our solution becomes the well-known periodic instanton.
In Sec.~VII, we take the zero temperature limit and obtain the
instanton solution in $R^4$. In Sec.~VIII, we find the moduli space
and its metric.  In Sec.~IX, we conclude with some remarks.

\section{The Nahm Construction}

The Nahm construction uses the Nahm data and the solution of the ADHMN
equations to construct the self-dual magnetic monopole
configurations~\cite{nahm2,adhm}. In addition, by studying the moduli
space of the Nahm data, one can construct the moduli space metric of
the corresponding magnetic monopole configurations. Especially in the
$SU(2)$ gauge theory, there has been considerable work in the Nahm
construction of magnetic monopoles~\cite{nahm2}.

For a $SU(2)$ gauge theory on $R^3\times S^1$, there are three relevant
time intervals for the Nahm equation,
\beq
-\frac{\pi}{\beta}<t<-\frac{u}{2},\;\; \,\,
-\frac{u}{2}<t<\frac{u}{2},\;\; \,\, \frac{u}{2}<t<\frac{\pi}{\beta}.
\eeq
Because we are
considering  calorons, we should require  the Nahm data to be periodic
in the the time variable $t$ in the Nahm equations~\cite{nahm}. The
first and  last intervals correspond to  monopoles of topological
charge $k_2$ and the second interval corresponds to monopoles of
topological charge $k_1$.  Since a single instanton is made of  two
distinct   monopoles, we need to introduce the jumping  condition at the
boundary $t=\pm u/2$. 

In general, the  Nahm data consist of a family of
triple hermitian matrix functions ${\bf T}(t)$, of dimensions
$l(t)\times l(t)$, defined in every interval, together with triple
matrices $\balpha_p$ of dimension $l(t_p)\times l(t_p)$, defined at
each point $t_p$ where $l(t)$ does not jump. The value $l(t)$ in each
interval is the number of corresponding monopoles.   These should satisfy
the Nahm equations
\begin{equation}
\frac{dT_i}{dt} - i[T_4,T_i] = -i\epsilon_{ijk}T_j T_k + \sum_p
(\balpha_P)_i  \delta(t-t_P) .
\label{nahm}
\end{equation}
When $l(t_P-\epsilon)\ne l(t_P+\epsilon)$, there is a usual boundary
condition on Nahm data on both sides of $t_P$. Since the time-interval
is periodic in $2\pi/\beta$, ${\bf T} (-\pi/\beta) = {\bf
T}(\pi/\beta)$. Associated with $\balpha_P$, there exists
$2l(t_P)$-component row vector $a_P$ satisfying
\beq
a_P^\dagger a_P = \balpha_P \cdot \bsigma - i (\alpha_P)_0
I_{2l(t_P)\times 2l(t_P)}
\label{jump1}
\eeq

The next step is to find a $2l(t) \times 2$ matrix functions $v(t)$
and $2$-component row vectors $S_p$ obeying the ADHMN  equations
\beq
0 = \biggl[ -\frac{d}{dt} + ({\bf T} + {\bf x} )\cdot\bsigma + ix_4
\biggr] v  + \sum_P a_P^\dagger S_P \delta(t-t_P)
\label{adhmn}  
\eeq
The matrices $v(t)$ is periodic, $v(-\pi/\beta) = v(\pi/\beta)$. 
These matrices should satisfy the normalization condition
\beq 
I_{2\times 2} = \int_{-\pi/\beta}^{\pi/\beta} dt\, v^\dagger v +
\sum_P S_P^\dagger S_P 
\label{norm1}
\eeq
Then, the corresponding  self-dual gauge field configuration is 
given by 
\beqn
A_\mu &=& \int^{\pi/\beta}_{-\pi/\beta} dt \, v^\dagger (t) \partial_\mu
v(t) + \sum_P  S_P^\dagger \partial_\mu S_P \nonumber \\
&=& \frac{1}{2} \int^{\pi/\beta}_{-\pi/\beta} dt \, [v^\dagger (t)
\partial_\mu v(t)  - \partial_\mu v^\dagger (t)  v(t) ] \\ 
\nonumber
& & \; + \frac{1}{2}\sum_P [ S_P^\dagger \partial_\mu S_P- 
  \partial_\mu S_P^\dagger  S_P]
\label{gauge1}
\eeqn

In this paper we will concern with the field configuration for two
distinct fundamental monopoles, so that $l(t)=1$ for the entire
interval $ [ -\pi/\beta, \pi/\beta] $.  The solutions of the Nahm
equations at each interval are  trivial.  We rotate and translate the
field configuration so that two massive monopoles lie on the
$z$-axis. The corresponding Nahm data is
\beqn
& & \bT_0 = \bT_2  = -\bx_2 = -(0,0,(\bx_{2})_3) \nonumber \\
& & \bT_1 = - \bx_1= - (0,0,(\bx_{1})_3) 
\eeqn
where $\bx_1, \bx_2$ are the positions of two massive monopoles.  In
our choice, the distance between two monopoles is $D=(\bx_{2}-\bx_{1})_3
>0$. For a given coordinate point $\bx$, we introduce its relative
positions with respect to two 
monopoles, as shown in Fig.I,
\beq
\by_1 = \bx - \bx_1,\,\,\, \by_2 =\bx-\bx_2,
\eeq
and  weighted relative positions 
\beq
\bs_1 = u\by_1,\,\,\, \bs_2=(\frac{2\pi}{\beta}-u)\by_2.
\eeq
The center-of-mass position is
\beq
\bx_{\rm cm} = k_1 \bx_1 + k_2\bx_2.
\label{cm}
\eeq

\begin{center}
\leavevmode
\epsfxsize=2.0in
\epsfbox{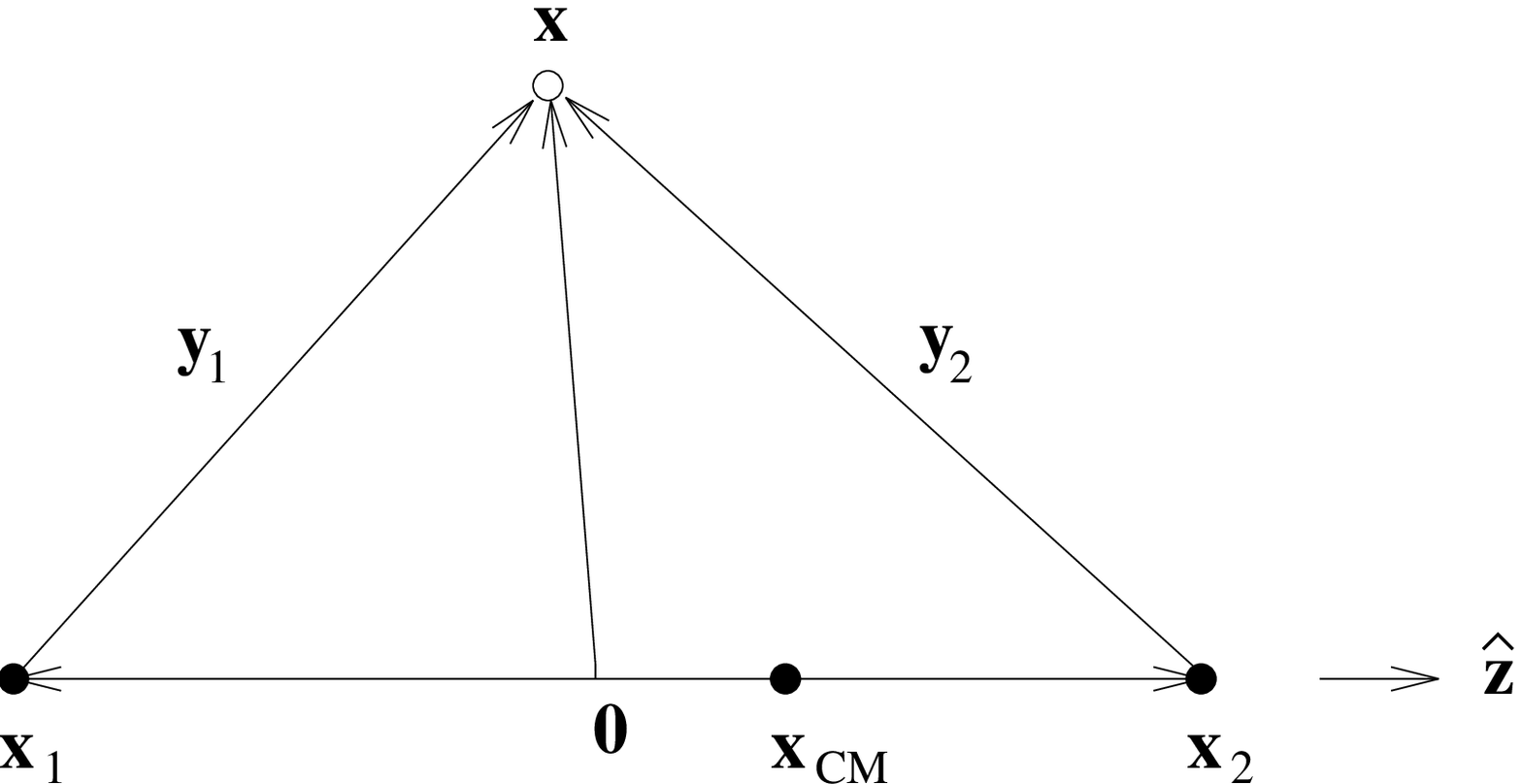}
\end{center}
\centerline{{\bf Figure I}: The position vectors for two magnetic monopoles.}
\vspace{0.5cm}

{}From Eq.~(\ref{nahm}) we get the jumping functions $(\balpha_1)_i =
D\delta_{i3}$ and $(\balpha_2)_i = -D\delta_{i3}$. Their 
corresponding two row vectors in Eq.(\ref{jump1}) are then
\beq
 a_1 = (\sqrt{2D},0), \,\, a_2 = (0,\sqrt{2D}) .
\eeq
The solutions of the ADHMN equation (\ref{adhmn}) at each interval can
be expressed as  
\beqn & & v_0(t)= \frac{1}{\sqrt{N_2}} e^{[ix_4 + \sigma \cdot \by_2
](t+\pi/\beta)} \,C_2\,\,\,\,\,\,\,\, {\rm for}\,\, t \in
[-\frac{\pi}{\beta},-\frac{u}{2}), \nonumber \\ 
& & v_1(t)= \frac{1}{\sqrt{N_1}}
e^{[ix_4 + \sigma \cdot \by_1 ]t}\,C_1 \;\;\;\;\;\;\;\;\;\;\;\;\;
 {\rm for}\,\,t\in
(-\frac{u}{2},\frac{u}{2}), \nonumber \\
& & v_2(t) = \frac{1}{\sqrt{N_2}}
e^{[ix_4 + \sigma \cdot \by_2 ](t-\pi/\beta)}\, C_2 \,\,\,\,\,\;\;
{\rm for}\,\, t\in (\frac{u}{2}, \frac{\pi}{\beta}],
\label{vss}
\eeqn
where $C_i$ are $2\times 2$ matrices  and
\beq
 N_i= \frac{1}{y_i}\sinh s_i\,\,(i=1,2).
\eeq
The  periodic condition $v_0(-T/2)=v_2(T/2)$ is automatically  satisfied. 
For this solution, the normalization condition (\ref{norm1}) becomes
\beq
I_{2\times 2} = C_1^\dagger C_1 + C_2^\dagger C_2 +
S_1^\dagger S_1 + S_2^\dagger S_2 .
\label{norm2}
\eeq

To find $C_1,C_2$ and $S_1, S_2$, we use the normalization condition
(\ref{norm2}) and the discontinuity equations derived from
Eq.~(\ref{adhmn}).  In addition, we require the gauge field
(\ref{gauge1}) to be  single-valued. Then, $C_i$ and $S_i$ are 
determined uniquely up to  acceptable gauge transformations. 
The $S_1$ and $S_2$  can be regarded as the first and second row
vectors of a $2\times 2$ matrix $S$, which takes the explicit form
\beq
 S = \frac{1}{\sqrt{{\cal N}}} e^{-i\frac{u}{2} x_4\sigma_3 },
\label{ess}
\eeq
where 
\beqn
{\cal  N} &=& 1 + \frac{2D}{{\cal M}}  \biggl\{  N_1 \bigl(\cosh s_2
-(\hat{\by}_{2})_3 \sinh s_2 \bigr) 
+ N_2 \bigl(\cosh s_1 +(\hat{\by}_1)_3 \sinh s_1
\bigr) \biggr\}.
\label{normal}
\eeqn
The two matrices $C_i$ are more complicated.  It is useful
first to introduce two $2\times 2$ matrices,
\beqn
& & B_1 = e^{i\frac{\pi}{\beta} x_4 } e^{-\frac{\sigma}{2}\cdot \bs_1}
e^{-\frac{\sigma}{2}\cdot\bs_2} 
- e^{-i\frac{\pi}{\beta} x_4/} e^{\frac{\sigma}{2}\cdot \bs_1}
e^{\frac{\sigma}{2}\cdot\bs_2} ,
\nonumber \\
& & B_2 =  e^{\frac{\pi}{\beta} ix_4 }e^{-\frac{\sigma}{2}\cdot \bs_2}
e^{-\frac{\sigma}{2}\cdot\bs_1} 
- e^{-i\frac{\pi}{\beta} x_4 } e^{\frac{\sigma}{2}\cdot \bs_2} 
e^{\frac{\sigma}{2}\cdot\bs_1} ,
\eeqn
and a  scalar quantity
\beq
 {\cal M} = 2\biggl\{ \cosh s_1 \cosh s_2
+\hat{\by}_1\cdot\hat{\by}_2  \sinh s_1  \sinh s_2   -\cos (
\frac{2\pi}{\beta} x_4 ) \biggr\},  \\
\eeq
where  ${\cal M} = B_1 B_1^\dagger =B_2 B_2^\dagger$.
Then the desired expression for $C_i$'s are given as 
\beqn
 C_1&=&  \sqrt{\frac{2DN_1}{{\cal N}}} \frac{B_1^\dagger}{{\cal M}}
\biggl[ e^{-\frac{\sigma}{2}\cdot\bs_2} Q_+ + e^{\frac{\sigma}{2}\cdot\bs_2}
Q_- \biggr] e^{-i\frac{\pi}{\beta} x_4 \sigma_3}, \nonumber  \\
 C_2 &=&  \sqrt{\frac{2DN_2}{{\cal N}}} \frac{B_2^\dagger}{{\cal M}}
\biggl[ e^{\frac{\sigma}{2}\cdot\bs_1}Q_+ +
e^{-\frac{\sigma}{2}\cdot\bs_1} Q_- \biggr] 
\eeqn
with projection operators 
\beq
Q_\pm = \frac{1\pm \sigma_3}{2}.
\eeq

The gauge field (\ref{gauge1}) becomes
\beqn
A_\mu(\bx,x_4) = & & C_1^\dagger V_\mu({\by_1};u) C_1 + C_2^\dagger 
V_\mu({\by_2};\frac{2\pi}{\beta}-u) C_2   \nonumber   \\
& &  + C_1^\dagger \partial_\mu C_1 +
C_2^\dagger \partial_\mu C_2 +S^\dagger
\partial_\mu S ,
\label{solution}
\eeqn
where $V_\mu(\bx;u)$ is the ordinary BPS monopole solution, 
\beqn
V_4(\bx;u) &=& \frac{\sigma_a}{2i}\hat{x}_a \biggl[\frac{1}{|\bx|} -
\frac{u}{\coth 
(u|\bx|)}  \biggr] , \nonumber \\
V_i(\bx;u) &=& \frac{\sigma_a}{2i} \epsilon_{aij} \hat{x}_j
\biggl[\frac{1}{|\bx|}- \frac{u}{\sinh (u |\bx|)} \biggr].
\eeqn
The field configuration (\ref{solution}) is the desired expression for
a single instanton. Under the gauge transformation $A_\mu \rightarrow
UA_\mu U^\dagger -\partial_\mu U U^\dagger $, we see $C_i\rightarrow
C_i U^\dagger$ and $S\rightarrow S U^\dagger$.

Notice  that ${\cal M}$  vanishes  at only one  point 
\beq
x_{\rm singular} = (\bx_{\rm cm}, x_4=0).
\label{singular1}
\eeq
The gauge field (\ref{solution}) turns out to have a gauge singularity
at this point as we will see later.

\section{Near Each Monopole}

To see the field configuration (\ref{solution}) describes two magnetic
monopoles of opposite charge, let us consider the limit
$D>>1/u,(2\pi/\beta-u)^{-1} $
and so their cores do not overlap. We expect the configuration to
approach that of each monopole near $\bx_1$ or $\bx_2$. If we are near
the first monopole so that  $|\by_1 |<< D$, we see easily that  
\beq
C_2, S_1, S_2 \sim\frac{1}{\sqrt{D}},
\eeq
and 
\beq
C_1 = \frac{\sigma_3 \cosh \frac{s_1}{2} - \bsigma\cdot \hat{\by}_1 \sinh
\frac{s_1}{2} }{\sqrt{\cosh s_1 -(\hat{\by}_1)_3 \sinh s_1}} + {\cal
O}(\frac{1}{D}) ,
\eeq
which is a single-valued unitary matrix. Thus, the whole gauge
configuration (\ref{solution}) becomes approximately a gauge
transformation of the single monopole configuration $V_\mu(\by_1;u)$.

Similarly, near the second monopole, $|\by_2|<< D$, we see that 
\beq
C_1,S_1,S_2 \sim \frac{1}{\sqrt{D}},
\eeq
and 
\beq
C_2 = \frac{-\sigma_3 \cosh \frac{s_2}{2} - \bsigma\cdot \hat{\by}_2 \sinh
\frac{s_2}{2} }{\sqrt{\cosh s_2 +(\hat{\by}_2)_3 \sinh s_2}}
e^{-i\frac{\pi x_3}{\beta}\sigma_3},
\eeq
which is an unitary matrix. Thus, the field configuration
(\ref{solution}) becomes a gauge transformation of the second monopole
configuration, but the sign of magnetic charge is changed by the large gauge
transformation $e^{-i \frac{\pi x_3}{\beta}\sigma_3}$.  The above
discussion shows that one can identify individual magnetic monopoles
when their cores are not overlapping.

\section{Outside Monopole Core}

Outside monopole core $s_1,s_2>>1$, we can neglect exponentially small
terms. Especially we see
\beqn
{\cal M} \approx \frac{1}{2}e^{ s_1 + s_2} \biggl(1 + \hat{\by}_1\cdot
\hat{\by}_2 \biggr), \nonumber \\
{\cal N} \approx 1 + \frac{D}{y_1 y_2} \frac{ y_1+y_2 + D}{1+
\hat{\by}_1 \cdot \hat{\by}_2 } .
\eeqn
{}From this we get 
\beqn
& C_1 & \approx \sqrt{\frac{D}{y_1 {\cal N}} } \frac{2}{1 +
\hat{\by}_1\cdot \hat{\by}_2} \biggl( P_{1-}P_{2-} Q_+ - P_{1+}P_{2+}
Q_- \biggr)  \\
& C_2 & \approx \sqrt{\frac{D}{y_2 {\cal N}}}  \frac{2}{1 +
\hat{\by}_1\cdot \hat{\by}_2} \biggl( P_{2-}P_{1-}Q_- - P_{2+} P_{1+}
Q_+ \biggl)  e^{-i \frac{\pi}{\beta} x_4  \sigma_3} 
\eeqn
where 
\beq 
  P_{i\pm} = \frac{1\pm \hat{\by}_i\cdot \bsigma}{2} \,\,\, (i=1,2)
\eeq
are projection operators.  Using these approximations, we can obtain
the field configuration outside the monopole core region. We expect
this to be purely Abelian and so a simple superposition of Abelian
fields in a unitary gauge. However seeing this explicitly does not
seem to be simple.

There are still immediate informations following the above
expressions.  For $s_1,s_2>>D$, $N\approx 1$ and $C_1\sim C_2\sim
\sqrt{D/|\bx|}$, making
\beq
A_\mu = -i\frac{u}{2} \sigma_3 \delta_{\mu 4}+ {\cal
O}\biggl(\frac{D}{|\bx|^2}\biggr) 
\eeq
This implies that when the distance between two monopoles goes to
zero, the field configuration becomes trivial, which is exactly what
we hope for the zero size instanton. Also the gauge field approaches
the vacuum trivially at spatial infinity, implying no boundary
contribution from spatial infinity to the topological charge
(\ref{top}). We will see in a moment that the only nontrivial
contribution comes from the  singularity~(\ref{singular1}).

\section{Near Singularity}

To consider the singularity at (\ref{singular1}), we put  the
center-of-mass at  the origin, so that $(\bx_1)_3=-k_2D$ and
$(\bx_2)_3=k_1 D$. Then, by expanding the matrices around $x_\mu = 0$,
we get
\beq
C_1\approx C_2 \approx  \frac{i}{\sqrt{2}} U(x)_s , \,\,S\approx {\cal
O}(1)
\eeq
where 
\beq
U(x)^\dagger_s =  \frac{x_4 + i \sigma_3 x_3 + i(\sigma_1x_1 + \sigma_2 x_2)
q}{\sqrt{ x_4^2 + x_3^2 + (x_1^2 +x_2^2)q^2} } 
\eeq
with $
q = \sinh (2\pi k_1 k_2 D/\beta)/ (2\pi k_1 k_2 D/\beta) $.

Thus the gauge field near the singularity $x_\mu=0$ becomes 
\beq
A_\mu = U_s \partial_\mu U_s^\dagger + {\cal O}(1) 
\eeq
showing that it is pure gauge singularity.  The nontrivial
contribution at this gauge singularity to the topological charge
(\ref{top}) is one, as expected for a single instanton.

\section{Massless monopole limit}

We choose $u=2\pi/\beta$. In this case the Wilson loop
becomes trivial. The gauge symmetry is restored to the original
$SU(2)$~\cite{caloron}.  In this limit the isolated second monopole solution
disappears as $V_\mu(y_2,2\pi/\beta-u)=0$. The size of the second monopole
becomes infinite and its topological charge vanishes. It loses
its meaning as an isolated object.

We put the massive monopole at the origin so that $\by_1 = \bx$ and
$\by_2= D\hat{\bz}$.  In this limit, $N_2=C_2=0$. After a
large gauge transformation, $e^{-i\frac{\pi}{\beta} x_4 \sigma_3}$,
one can see that the solution (\ref{solution}) is 
\beq
A_\mu = \frac{i}{2}\bar{\sigma}_{\mu\nu} \partial_\nu \ln {\cal N}
\eeq
with $\bar{\sigma}_{ij}=\epsilon_{ijk}\sigma_k$ and
$\bar{\sigma}_{i4}=-\sigma_i$. In this limit  the normalization
coefficient (\ref{normal}) is 
\beq
{\cal N } = 1 + \frac{D}{|\bx|} \frac{\sinh (\frac{2\pi}{\beta} |\bx
|) }{\cosh (\frac{2\pi}{\beta} |\bx|) -\cos( Tx_4)} .
\eeq
This is exactly the periodic instanton solution~\cite{caloron}, once
we require a relation
\beq
D = \frac{\pi \rho^2}{\beta}
\label{rho}
\eeq
between the inter-monopole distance $D$ and the instanton scale
parameter $\rho$. In the zero temperature limit, $\beta\rightarrow
\infty$, one can see that finite size instanton solution can be
obtained only if the distance between two magnetic monopoles
approaches zero.

The interpretation of this solution can be done consistently with the
previous pictures about massless monopole~\cite{lwy3,piljin}.  First of
all, when we remove the massless monopole, $D\rightarrow \infty$, the 
configuration becomes pure magnetic monopole~\cite{rossi}.
When the massless monopole is at finite distance, the field
configuration near the massive monopole is purely magnetic and then
the massless monopole or the nonabelian cloud shields the magnetic
charge of the massive monopole at distance scale $D$ and the field
configuration at scale $r>> D$  falls off
quickly like a dipole field configuration~\cite{pisarski}.

\section{zero temperature limit}

Let us now investigate our solution at  the zero temperature
limit $\beta\rightarrow \infty$, which implies $u\rightarrow 0$ by
Eq.~(\ref{weyl}). After putting the center-of-mass position (\ref{cm})
at the origin,  we see that for finite $x = (\bx, x_4)$, $N_1 \approx
u$, $N_2 \approx 2\pi/\beta - u $, and
\beqn
{\cal M} \approx (2\pi/\beta^2)^2 x^2  \nonumber \\
{\cal N} \approx 1 + \frac{\beta  D}{\pi  x^2} 
\eeqn
Thus the zero temperature limit of $S$ in Eq.~(\ref{ess}) is
nontrivial only if $\beta D$ remains finite. This is consistent with
the argument after Eq.~(\ref{rho}).  After removing the
singularity at the origin by a singular gauge transformation,
$U^\dagger = (x_4 +i \bsigma\cdot\bx) /\sqrt{x^2}$, 
a  $2\times 2$ 
matrix $S$ of Eq.(\ref{ess}) 
becomes
\beq
S = \frac{x_4 + i \bsigma\cdot \bx}{\sqrt{x^2 + \rho^2} }
\eeq
with $\rho^2=\beta D/\pi$ as shown in Eq.~(\ref{rho}). 
The two matrices $v_1,v_2$ of Eq.~(\ref{vss}) are simply
\beq
v_1\approx v_2\approx -\frac{\beta}{2\pi} \sqrt{\frac{2D}{x^2}}
\eeq
The gauge field ({\ref{gauge1}) becomes
\beq
A_\mu = \frac{-i\sigma_{\mu\nu}x_\nu}{x^2+\rho^2}
\label{regular}
\eeq
where $\sigma_{ij}=\epsilon_{ijk}\sigma_k$ and $\sigma_{i4}=\sigma_i$.
This is the standard regular expression for a single instanton on
$R^4$~\cite{bpst}.

\section{Moduli Space Metric}

The relative moduli space of two constituent monopoles for a single
instanton is known to be the Taub-NUT space with $Z_2$
division~\cite{piljin}. Here we fix the normalization and provide the
global picture of the moduli space, which also shed light on the zero
temperature limit and the trivial Wilson loop limit.

To fix the normalization, we consider the additional real time
direction $x^0$, which makes our theory to be five
dimensional. Instantons and magnetic monopoles appear as self-dual
solitons.  The number of zero modes of a single instanton is eight and
is the sum of the zero modes for constituent monopoles. Each monopole
carries four zero modes for its position and internal $U(1)$ phase.
We can divide eight instanton zero modes into four for the center of
mass motion of monopoles and four for the relative motions of magnetic
monopoles. Defining the moduli space is quite similar to the monopole
case~\cite{gauntlett}. For the infinitesimal change of the moduli
parameters $z_A$, $A=1,..,8$, the corresponding infinitesimal change
$\delta_A A_\mu$ would satisfy the background gauge and the linearized
self-dual equations. Then the moduli space metric is given by
\beq
{\cal G}_{AB} = \int d^4x\, \delta_A A_\mu \delta_B A_\mu
\eeq
One can easily see that this space should be hyper-K\"ahler by
generalizing the argument in Ref.~\cite{gauntlett}.

The detailed derivation of the moduli space of these monopoles are given
before~\cite{lwy2,lwy1}. (Since their magnetic charges belong to the
same $U(1)$ group with opposite sign, the value of the parameter
$\lambda $ in Ref.~\cite{lwy2} is two.)  Each monopoles are imagined
to carry corresponding integer quantized electric charge, $q_1,q_2$.
the only modification for the case in hand is that we have to integrate over
$x_4$.  This leads to an overall multiplicative factor $\beta$ on the
effective low energy Lagrangian. The center-of-mass moduli space is
just $R^3\times S^1$.  Since $\beta(m_1+m_2) =8\pi^2$, the metric for
the center-of-mass moduli space becomes
\beq
ds^2_{\rm cm} = 8\pi^2 (d{\bf R}^2 + \frac{\beta^2}{4\pi^2} d\chi^2)
\label{center}
\eeq
where ${\bf R}$ is the center-of-mass position and $\chi$ is the
conjugate variable for the total electric charge.  The total electric
charge is $q_\chi=k_1q_1 + k_2 q_2$~\cite{lwy2}, which turns out to be the
$x_4$ momentum~\cite{piljin}.  This charge needs not to be
quantized~\cite{piljin} and $\chi$ lies along the real line $R$.
Thus, we cannot identify $\beta \chi/(2\pi)$ with $x_4$, unless
$q_1=q_2$. The overall coefficient $8\pi^2$ is the mass of
instanton.

The relative mass $m_1m_2/(m_1+m_2)$ between two monopoles is $8\pi^2
k_1 k_2/\beta$. We introduce the relative position between two
monopoles ${\bf r} = \bx_1-\bx_2$ and note that $|\br| =D$.  The
metric for the relative moduli space (obtained after multiplying $\beta$
to Eq.(5.8) in Ref.~\cite{lwy2}.)  is 
\beq
ds^2_{\rm rel} = 8\pi^2 k_1 k_2  \biggl[
(1 + r_0/r)d{\bf r}^2 + r_0^2(1+r_0/r)^{-1} (d\psi + {\bf w}({\bf
r})\cdot d{\bf r})^2\biggr]
\label{rel}
\eeq
where $r_0 = \beta/(2\pi k_1k_2)$ and ${\bf w}({\bf r})$ is the Dirac
potential such that $\nabla\times {\bf w} = \nabla (1/r)$. This is the
Taub-NUT space with length paramter $r_0/2$.  Since both monopoles can
carry only integer electric charge, their relative charge $q_\psi
=q_1=q_2$ is integer quantized instead of half-integer quantized as
in the SU(3) case~\cite{lwy1}.
Thus their relative phase $\psi$ should have the interval $[0,2\pi]$
instead of $[0,4\pi]$. This is the origin of $Z_2$ orbifold
singularity of the relative moduli space ${\cal M}_0$. 
 The total moduli space can be found by a similar
discussion as for monopoles~\cite{lwy2,lwy1} and is given as 
\beq
{\cal M} = R^3 \times \frac{R^1\times {\cal M}_0}{Z},
\eeq
where the generator of the identity map $Z$ is  $(\chi,\psi)=
(\chi+2\pi, \psi + 2\pi k_2)$.

In the zero temperature limit $\beta\rightarrow \infty$, or in the
limit where symmetry is restored, say, $k_2 \rightarrow 0$, the
relative metric becomes flat. This is similar to the massless limit of
the relative moduli space metric in $SO(5)$~\cite{lwy3}. After using
the instanton scale parameter $\rho$ in Eq.~(\ref{rho}), the metric
(\ref{rel}) becomes
\beq
ds^2 = 16\pi^2(d\rho^2 + \rho^2 d \Omega_3^2)
\eeq
where $d\Omega^2_3$ is the metric of a unit three sphere.  The overall
coefficient can be checked directly by calculating $\int d^4 x
(\delta_\rho A^a_\mu)^2 $, which is straightforward because $\partial
A_\mu^a /\partial \rho$ of Eq.~(\ref{regular}) satisfies the
background gauge, $D_\mu \delta A_\mu = 0$. Since the adjoint
matters belongs to $SO(3)$,  so the gauge orbit of a single
instanton  is $S^3/Z_2$, implying the $Z_2$ orbifold singularity
at origin.

\section{Concluding Remarks}

By using the Nahm construction, we have found the field configuration
for a single instanton in the $SU(2)$ gauge theory on $R^3\times
S^1$. When the gauge group is spontaneously broken by the Wilson loop,
a single instanton is shown to be composed of two fundamental
monopoles of opposite magnetic charge. By taking various limits, our
solution is shown to be consistent with the previously known ideas
about periodic instantons, massless monopoles and zero temperature
instantons.

There are several interesting implications from our work as mentioned
in Ref.~\cite{piljin,kimyeong}. Here we also see that the zero temperature
limit may  be interesting.  At zero temperature limit of a single
caloron, the positions of both two monopoles should come together to
the center in order to get a finite size instanton, which  makes
the monopole picture somewhat trivial. However the story cannot be
all there is for the two caloron case. Even at the zero temperature limit of
two close-by calorons, there are no identifiable instanton
positions~\cite{jackiw}. Thus, it is not clear where the four
constituent monopoles for two calorons will end up at the zero
temperature limit.  Thus, we hope that the picture of composite
instantons and their constituent monopoles still survives even at zero
temperature in some sense, say, after abelian
projection~\cite{thooft}, and leads to new insight on understanding
the chiral symmetry and confinement in zero temperature QCD.

\centerline{\bf Acknowledgments} 

We like to thank Erick Weinberg and Piljin Yi for useful discussions
on the Nahm construction.  This work is supported in part by the
U.S. Department of Energy.

\vspace{5mm}

While writing up this paper, we become aware of Ref.~\cite{kraan}
which has a considerable overlap with our work.

\end{document}